# A combined thermal-resistance-capacity and finite-element model for very fast and accurate short- and medium-term simulations of single U-tube borehole heat exchangers

Enzo Zanchini [a*], Francesco Zanchini [b], Claudia Naldi [a]

[a] Department of Industrial Engineering, Viale del Risorgimento 2,
University of Bologna, Bologna, Italy

[b] Student of the Department of Physics and Astronomy, Viale Carlo Berti Pichat 6/2,
University of Bologna, Bologna, Italy

[*]Corresponding author - Email: enzo.zanchini@unibo.it

**Abstract**

An accurate design of a ground-coupled heat pump system requires the knowledge of the outlet fluid temperature from the borehole heat exchangers (BHEs), both in the short and long term. This paper fucuses on the short and medium term. In this time range, either 3D finite-element simulations or Thermal Resistance Capacity Models (TRCMs) can be applied. The former can yield very accurate results but require long computation times. The latter are much faster but cannot be fully precise, because they require simplifying assumptions. In this paper, we present a new method for the short-term and medium-term simulation of single U-tube BHEs, which combines the speed of TRCMs and the accuracy of finite-elements simulations. The method uses a TRCM to estimate the thermal response of the BHE, then corrects the results by interpolation with a dataset, which has been produced by running 54 finite-element simulations in various configurations. The model is implemented in a C++ program, available at the open-source online data repository of the University of Bologna. The program provides, within two seconds, the time evolution of the inlet, outlet and mean fluid temperature, of the mean BHE surface temperature, of the 3D and the effective borehole thermal resistance. It can be easily connected to long-term simulation tools to obtain the full-time-scale thermal response of a bore field.

**Nomenclature**

| | |
|---|---|
| ANN | Artificial neural network |
| b | Borehole node |
| BHE | Borehole Heat Exchanger |
| $c$ | Specific heat capacity (J kg$^{-1}$K$^{-1}$) |
| $c_{\text{coeff}}$ | Correction coefficient, defined in Eq. (32) |
| $C$ | Heat capacity (J K$^{-1}$) |
| $D$ | Buried depth (m) |
| $f$ | Weighting factor, defined in Eq. (2) |
| $f_1, f_2$ | downward and upward fluid nodes |
| FLS | Finite line-source |
| $g_i$ | $i$-th ground node, $i = 1, 2, \ldots, n + 1$ |
| $gt_1, gt_2$ | grout nodes |
| $h$ | Convection coefficient (W m$^{-2}$K$^{-1}$) |
| HUST | Horizontally uniform surface temperature |
| $k$ | Thermal conductivity (W m$^{-1}$K$^{-1}$) |
| $l$ | Height of each horizontal slice (m) |
| $L$ | BHE height (m) |
| $m$ | Number of horizontal slices |
| $\dot{m}_f$ | Mass flow rate of the fluid (kg s$^{-1}$) |
| $n$ | Number of ground annuli |
| $Nu$ | Nusselt number |
| $p_1, p_2$ | pipe nodes |
| $\dot{q}_l$ | Mean heat rate per unit length (W m$^{-1}$) |
| $\dot{Q}$ | Heat rate supplied to the BHE (W) |
| $r_0, r_b$ | BHE radius (m) |
| $r_i$ | Radius of the outer surface of the $i$-th ground annulus (m) |
| $\bar{r}_i$ | Radius defined in Eq. (10) (m) |
| $r_{\text{pe}}$ | Outer radius of the pipe (m) |
| $r_{\text{pi}}$ | Inner radius of the pipe (m) |
| $r_x$ | Radial distance from the BHE axis, defined in (Ruiz-Calvo et al., 2015) (m) |
| $R_1^\Delta$ | Thermal resistance per unit length between fluid 1 and the BHE wall (m K W$^{-1}$) |
| $R_2^\Delta$ | Thermal resistance per unit length between fluid 2 and the BHE wall (m K W$^{-1}$) |
| $R_{12}^{\Delta}$ | Thermal resistance per unit length between fluid 1 and fluid 2 (m K W$^{-1}$) |
| $R_a$ | Total thermal resistance per unit length between the flows, (m K W$^{-1}$) |
| $R_b$ | 2D steady-state borehole thermal resistance per unit length (m K W$^{-1}$) |
| $R_{b3D}$ | 3D borehole thermal resistance per unit length (m K W$^{-1}$) |
| $R_{beff}$ | Effective borehole thermal resistance per unit length (m K W$^{-1}$) |
| $R_{conv}$ | Convective thermal resistance per unit length (m K W$^{-1}$) |
| $R_i$ | Thermal resistance per unit length of the ground between $\bar{r}_{i-1}$ and $\bar{r}_i$ (K W$^{-1}$) |
| $s$ | Half shank spacing (m) |
| $t$ | Time (s) |
| $T$ | Temperature (°C) |
| $T_b$ | Mean temperature of the BHE surface (°C) |
| $T_{fave}$ | $= (T_{in} + T_{out})/2$ (°C) |
| $T_{fm}$ | Mean temperature of the fluid (°C) |
| $T_g$ | Undisturbed ground temperature (°C) |
| $T_{in}$ | Inlet fluid temperature (°C) |
| $T_{out}$ | Outlet fluid temperature (°C) |
| TRCM | Thermal Resistance Capacity Model |
| $\dot{V}$ | Volume flow rate (m$^3$s$^{-1}$) |
| $\dot{V}_0$ | Reference volume flow rate (m$^3$s$^{-1}$) |

Subscripts

| | |
|---|---|
| 0 | Reference, initial, of the borehole node |
| c | Corrected |
| f | Of fluid |
| f1, f2 | Of the fluid node $f_1$, of the fluid node $f_2$ |
| g | Of ground |
| gt | Of grout |
| gt1, gt2 | Of the grout node $gt_1$, of the grout node $gt_2$ |
| i | Of the $i$-th ground annulus, of the $i$-th ground node, $i = 1, 2, \ldots, n+1$ |
| j | Of the $j$-th horizontal slice, $j = 1, 2, \ldots, m$ |
| k | Of the $k$-th time instant, $k = 1, 2, \ldots K$ |
| p | Of pipe |
| p1, p2 | Of the pipe node $p_1$, of the pipe node $p_2$ |

Greek symbols

| | |
|---|---|
| $\varepsilon$ | Fraction of the grout heat capacity placed in the grout nodes |
| $\eta$ | Dimensionless parameter, defined in Eq. (7) |
| $\rho$ | Density (kg m$^{-3}$) |
| $(\rho c)$ | Volumetric heat capacity (J m$^{-3}$K$^{-1}$) |
| $\varphi$ | Dimensionless coefficient, defined in Eq. (3) |

**1. Introduction**

Ground-coupled heat pumps are very efficient systems for heating and cooling buildings. Usually, the heat exchange with the ground is obtained by means of vertical ground heat exchangers, called Borehole Heat Exchangers (BHEs), composed of either a single U-tube or a double U-tube in high density polyethylene, surrounded by a sealing grout.

An accurate design of a ground-coupled heat pump system requires the knowledge of the time evolution of the fluid temperature at the outlet of the bore field, $T_{\text{out}}$, both in the short and long term. This thermal response is usually determined by employing a dimensionless function, called *g-function*, which yields the time evolution of the mean temperature of the surface between the BHEs and the ground, $T_{\text{b}}$, due to a time constant heat load. The time evolution of $T_{\text{b}}$ produced by a time dependent heat load is then obtained by the superposition of the effects in time.

The simplest way to determine the *g-function* of a bore field is to use the finite line-source (FLS) model with uniform heat load. This condition allows determining the *g-function* of the bore field using the analytical expression of the temperature field around a single BHE [1] and the superposition of the effects in space. However, the FLS model with uniform heat load yields an overestimation of the *g-function* of the bore field, that can be relevant for large and compact bore fields [2, 3, 4].

A more accurate FLS model is the one proposed by Eskilson [5], in which the condition of uniform heat load is replaced by that of uniform surface temperature of the bore field. A semi-analytical method to determine *g-functions* of bore fields with Eskilson's boundary condition was developed by Cimmino and Bernier [6], and numerical methods were proposed by Monzó et al. [7] and by Naldi and Zanchini [8].

The condition of uniform surface temperature yields a slight underestimation of the *g-function* of a bore field [9]. A semi-analytical model that employs the real condition of BHEs fed in parallel with the same inlet temperature was developed by Cimmino [9], and improved in subsequent papers [10, 11, 12]. Numerical models using the condition of uniform fluid temperature were proposed by Monzó et al. [13] and by Zanchini et al. [14], and a semi-analytical model with this condition was developed by Zanchini and Zanchini [15]. For U-tube BHEs, the condition of uniform fluid temperature yields the same *g-function* as the condition of BHEs fed in parallel with the same inlet temperature [16]. The time evolution of $T_{\text{out}}$ can then be obtained from that of $T_{\text{b}}$ either through the mean temperature of the fluid, $T_{\text{fm}}$, or through the arithmetic mean of inlet and outlet temperature, $T_{\text{fave}} = (T_{\text{in}} + T_{\text{out}})/2$.

The mean fluid temperature, $T_{\text{fm}}$, can be deduced from $T_{\text{b}}$ by the relation

$$T_{\text{fm}} = T_{\text{b}} + \dot{q}_l R_{\text{b3D}}, \tag{1}$$

where $\dot{q}_l$ is the mean heat rate per unit length supplied to the bore field, and $R_{\text{b3D}}$ is the 3D borehole thermal resistance. The outlet fluid temperature can then be obtained from $T_{\text{fm}}$ through the equation proposed by Beier and Spitler [17], namely

$$T_{\text{out}} = T_{\text{fm}} - f \frac{\dot{Q}}{\dot{m}_f c_f}, \tag{2}$$

where $\dot{Q}$ is the mean heat rate supplied to each BHE, $\dot{m}_f$ is the mass flow rate circulating in each BHEs, $c_f$ is the specific heat capacity at constant pressure of the fluid, and $f$ is a weighting factor that can be determined as described in [17]. As an alternative, $T_{\text{out}}$ can be obtained from $T_{\text{fm}}$ through the equation proposed by Jahanbin et al. [18], namely

$$T_{\text{out}} = T_{\text{fm}} - \left(0.5 - \varphi \frac{\dot{V}}{\dot{V}_0}\right) \frac{\dot{Q}}{\dot{m}_f c_f}, \tag{3}$$

where $\dot{V}$ is the fluid volume flow rate in each BHE, $\dot{V}_0$ is a reference volume flow rate, equal to 12 liters per minute, and $\varphi$ is a dimensionless coefficient that can be determined by the correlations given in Jahanbin et al. [18] for single U-tube BHEs and in Zanchini and Jahanbin [19] for double U-tube BHEs.

The time evolution of $T_{\text{fave}}$ can be deduced from that of $T_{\text{b}}$ by the relation

$$T_{\text{fave}} = T_{\text{b}} + \dot{q}_l R_{\text{beff}}, \tag{4}$$

where $R_{\text{beff}}$ is the effective borehole thermal resistance. The time evolution of $T_{\text{out}}$ can then be obtained from that of $T_{\text{fave}}$ by the equation

$$T_{\text{out}} = T_{\text{fave}} - 0.5 \frac{\dot{Q}}{\dot{m}_f c_f}. \tag{5}$$

When the heat transfer within each BHE is quasi-stationary, $R_{\text{b3D}}$ is usually assumed to be equal to the steady-state 2D thermal resistance of a BHE cross section, $R_{\text{b}}$, that can be determined accurately by analytical expressions based on the multipole method [20, 21]. $R_{\text{beff}}$ can then be calculated from $R_{\text{b}}$ using Hellström's method [22], recalled and extended to double U-tube BHEs by Claesson and Javed [21]. Indeed, the quasi-stationary value of $R_{\text{b3D}}$ is slightly greater than $R_{\text{b}}$ [16], and the quasi-stationary value of $R_{\text{beff}}$ is slightly greater than that given by Hellström's method, as is shown in this paper. However, the underestimation of $T_{\text{out}}$ caused by assuming $R_{\text{b3D}} = R_{\text{b}}$ or by applying Hellström's method to determine $R_{\text{beff}}$ can be considered as acceptable.

On the contrary, in transient conditions, namely soon after the start of the BHE operation or a sudden change in heat load, the time-dependent values of $R_{\text{b3D}}$ and of $R_{\text{beff}}$ defined by Eqs. (1) and (4) are much lower than the estimated quasi-stationary values. As a consequence, it is necessary to perform an accurate short-term simulation of the BHE, to obtain the correct time evolution of $T_{\text{out}}$.

The short-term simulation of a BHE can be performed using an analytical solution, a 3D numerical simulation, or a lumped-parameter model.

Analytical solutions of short-term BHE models where the fluid is replaced by a heat generating solid, or surface, or set of lines, have been proposed by Lamarche and Beauchamp [23], Bandyopadhyay et al. [24, 25], Man et al. [26], Li and Lay [27], Lamarche [28]. These models, even if accurate, cannot take into account the effects of the heat exchange between the descending and the ascending flow, which enhances $R_{\text{b3D}}$ and, as a consequence, $T_{\text{fm}}$ [16].

An analytical short-term model of single U-tube BHEs that considers the energy balances along the flow has been developed by Beier [29]. In the model, the pipes are replaced by two adjacent half pipes separated by a thermal resistance and surrounded by the grout. The solution is given in the Laplace transformed domain and requires a numerical inverse Laplace transform algorithm.

Numerical 3D simulations of U-tube BHEs considering the energy balance along the flow have been performed by several authors. Li and Zheng [30], Florides et al. [31], Ozudogru et al. [32], Lei et al. [33], and Zanchini [16] proposed models with 3D simulation of the solid domain and 1D energy balance along the flow. Marcotte and Pasquier [34] and Pasquier and Marcotte [35] employed models with the 3D simulation of both the solid and the fluid domain, with a uniform velocity of the fluid. A 3D simulation can yield very accurate results, but requires a long computation time, typically several hours for each simulation.

As an alternative to numerical simulations that solve the local energy balance equations, some authors developed lumped-parameters numerical models based on the electric analogy, where the BHE and the ground are represented by a grid of thermal resistances and heat capacities.

De Carli et al. [36] developed a 3D Capacity Resistance model where the BHE and the ground are divided into $m$ horizontal slices, and the ground is divided into $n$ coaxial cylindrical regions. The conduction heat transfer is assumed to occur only in the radial direction, and the energy transfer in the vertical direction is due to the fluid flow. While both thermal resistances and heat capacities are considered for the ground elements, the heat capacity of the borehole is neglected. Zarrella et al. [37] developed an improved model, in which the heat capacities of both the sealing grout and the heat carrier fluid are considered. The model refers to double U-tube BHEs. The grout is divided into a core region, between the pipes, and a shell region, from the pipes to the borehole wall.

Bauer et al. [38] developed 2D Thermal Resistance Capacity Models (TRCMs) for coaxial, single U-tube, and double U-tube BHEs. The model for single U-tube BHEs is composed of two fluid nodes without heat capacities, two grout nodes, each with heat capacity equal to one half of the heat capacity of the grout, one borehole node without heat capacity, and several ground nodes. The thermal short circuiting between the pipes is considered by means of a grout-to-grout thermal resistance.

Bauer et al. [39] developed a 3D TRCM for single U-tube BHEs based on the 2D horizontal thermal circuit proposed in Bauer et al. [38]. In the new 3D model, the fluid nodes have the heat capacity of the corresponding fluid, and the horizontal thermal circuits are connected in the vertical direction by considering the heat transfer between overlapping nodes and the mass transport in the fluid. A validation of the model revealed a fair agreement between the outlet temperature of the fluid yielded by the model and that yielded by a finite-element simulation.

Pasquier and Marcotte [40] proposed an improvement of the 2D model by Bauer et al. [38], where each node is split in several nodes, and the grout thermal resistance and the grout-to-grout thermal resistance are spit in several equal parts in series. The same authors [35] transformed the 2D model of [40] into a quasi 3D model considering the vertical energy transfer by the fluid, and validated the new model by comparing its outcomes with those of a 3D finite-element model and with the experimental results of two thermal response tests.

Ruiz-Calvo et al. [41] developed a lumped-parameter model called Borehole-to-Ground where a vertical discretization of the BHE and the soil is introduced, and five nodes are considered for each depth: two fluid nodes, two grout nodes and one ground node. Each grout node has half of the heat capacity of the grout and is placed at a radial distance $r_x$ from the BHE axis, to be optimized. Each fluid node is connected to the ground node by four thermal resistances in series, whose expressions are given by Eqs. (13), (16), (17), and (21) of [41]. The thermal interference between the descending and the ascending flow is considered by means of a pipe-to-pipe and a grout-to-grout thermal resistance, given by Eqs. (18) and (19) of [41]. The model has no borehole node. This way, the authors avoided the assumption of an isothermal surface of the BHE, but were forced to use approximate expressions for the thermal resistances. Since the model employs only one ground node, it becomes inaccurate when the simulation period exceeds 18 hours.

Pasquier et al. [42] developed an artificial neural network (ANN) that constructs very rapidly both the short-term and the long-term thermal response of a field of single U-tube BHEs. The short-term thermal response is based upon a training set of 15,000 simulations performed with the TRCM of [35]. The ANN allows only one value of the inner and of the outer radius of the pipes and considers a vanishing convective thermal resistance between fluid and pipes. The MATLAB code of the ANN is available as supplementary material of the paper. Pasquier and Marcotte [43] extended the ANN to other values of the inner and outer radius of the pipes and to a larger set of allowed input parameters, still with a vanishing convective thermal resistance between fluid and pipes.

Most TRCMs assume a horizontally uniform surface temperature (HUST) of the BHE [35 – 40, 42, 43]. This assumption is necessary in order to use the accurate analytical expressions of the thermal resistances between each pipe and the borehole wall and between the pipes determined by the multipole method [22, 20]. However, as we will show in Section 2, the HUST condition yields an underestimation of $R_{b3D}$ which, in turn, yields an underestimation of $T_{fm}$, $T_{in}$, and $T_{out}$. The underestimation becomes appreciable after a few hours of operation.

In this paper we propose a new method for the short-term and medium-term simulation of single U-tube BHEs. The method is based on a TRCM, where the BHE and the ground are divided into horizontal slices, each modeled with a thermal circuit. The inaccuracies of the TRCM are corrected through a dataset of correction factors. The dataset has been produced by comparing the time evolution of $R_{b3D}$ yielded by the TRCM with that obtained by a very accurate 3D finite-element simulation in 54 different configurations. The correction factors in the remaining configurations are obtained by applying polynomial interpolation techniques to the dataset. The method has been implemented in a C++ program that yields, within two seconds, the same time evolution of $T_b$, $T_{fm}$, $T_{in}$, $T_{out}$, $R_{b3D}$, and $R_{beff}$ that can be obtained by a finite-element simulation requiring several computation hours. The C++ program is available at the open-source online data repository of the University of Bologna.

The paper is organized as follows. In Section 2, we describe our 3D finite-element simulation model and we use it to show that the HUST condition yields an underestimation of the thermal response. In Section 3, we describe the proposed TRCM. In Section 4, we compare the results of the TRCM with those of the finite-element model, we highlight the inaccuracies of the TRCM, and we point out that they can only be corrected a posteriori. In Section 5, we show that the inaccuracies can be resolved by introducing a time-dependent correction factor. In Section 6, we illustrate the results of the corrected TRCM.

## 2. Description of the finite-element model and effect of the HUST condition

The 3D finite-element simulation model employed in [16] was selected for the 3D simulations of this paper. The model was chosen due to its accuracy, demonstrated in [16] through mesh-independence tests and through the comparison with the FLS solution of Claesson and Javed [1] in a special case.

The model employs the Heat Transfer in Solids interface of COMSOL Multiphysics to simulate the heat conduction in the solid domains, and the COMSOL Pipe Flow Module to simulate the energy balance along the flow. The module is implemented with the method recommended in [16].

The fluid, represented by a line, flows in a closed circuit, and a heat generation is placed in the upper connection between the flows. For all simulations, the fluid considered is water, with thermal properties at 20 °C, taken from the NIST Chemistry WebBook [44]: density 998.21 kg/m$^3$, dynamic viscosity 1.0016 mPa s, thermal conductivity 0.59801 W/(m K), specific heat capacity at constant pressure 4.1841 kJ/(kg K). The convection coefficient, $h$, is determined through the Nusselt number, $Nu$, calculated by the Churchill correlation [45]. The pipes are represented by solid cylinders having the same thermal conductivity as the real pipes, and a reduced volumetric heat capacity, given by $0.5(\rho c)_\text{p}(r_\text{pe}^2 - r_\text{pi}^2)/r_\text{pe}^2$, as in [16]. This is to eliminate the overestimation of the heat capacity of the pipes due to considering solid pipes instead of the real annular ones.

In all simulations, the BHE is interred at a buried depth $D = 1.8$ m. The selected computational domain is a cylinder with radius 100 m and height 101.8 m greater than the BHE length, with the vertical axis oriented upwards and origin at the BHE top. The ground surface is considered as isothermal, with temperature $T_\text{g} = 0$ °C, equal to the undisturbed ground temperature and to the initial temperature. The vertical and the bottom surfaces of the computational domain are set to adiabatic. Although the radius of the computational domain could have been reduced, the same radius as in [16] was preferred, to use the meshes adopted in [16] as a reference.

To ensure high accuracy, the results are collected starting from $10^{-4}$ hours, the initial step is set to $10^{-6}$ hours, and the absolute accuracy is set to $10^{-4}$. To improve meshing, the lengths in the vertical direction are reduced by a scaling factor 20, and the thermal conductivities of the solids in the vertical direction are reduced by a scaling factor 400. The thermal conductivity of water, which is a scalar in the Pipe Flow Module, is reduced by the factor 400, and the Nusselt number is multiplied by 400.

The mesh used for the simulations is intermediate between Mesh 1 and Mesh 2 employed in [16]. This intermediate mesh, which will be denoted by Mesh 3, provides almost identical results to Mesh 2, but allows for a reduction in computation time. It is obtained by setting minimum element size 2 mm, maximum element size 4 m, element growth rate 1.25, curvature factor 1.

The U-tube BHE considered in this section will be denoted by BHE 1. It has length $L = 100$ m, radius $r_\text{b} = 76$ mm, half shank spacing $s = 47$ mm, and is placed in a ground with thermal conductivity $k_\text{g} = 1.8$ W/(m K). The volume flow rate of the fluid is 14 liters per minute, and the heat rate released to the fluid is $\dot{q}_l = 50$ W/m. The parameters that characterize BHE 1 are summarized in Table 1, fourth column. For BHE 1, one has $Nu = 80.2108$, $h = 1471.4$ W/(m$^2$K).

Table 1: Geometrical and thermal parameters of the BHEs employed in the examples

| Quantity | Symbol | Unit | BHE 1 | BHE 2 | BHE3 |
|---|---|---|---|---|---|
| BHE length | $L$ | m | 100 | 200 | 80 |
| BHE radius | $r_b$ | mm | 76 | 74 | 64 |
| Buried depth | $D$ | m | 1.8 | 1.8 | 1.8 |
| Half shank spacing | $s$ | mm | 47 | 40 | 32 |
| Pipe outer radius | $r_{pe}$ | mm | 20 | 20 | 16 |
| Pipe inner radius | $r_{pi}$ | mm | 16.3 | 16.3 | 13 |
| Pipe thermal conductivity | $k_p$ | W/(m K) | 0.4 | 0.4 | 0.4 |
| Pipe volumetric heat capacity | $(\rho c)_p$ | MJ/(m³K) | 1.824 | 1.824 | 1.824 |
| Grout thermal conductivity | $k_{gt}$ | W/(m K) | 1.0 | 1.8 | 2.0 |
| Grout volumetric heat capacity | $(\rho c)_{gt}$ | MJ/(m³K) | 2.5 | 2.3 | 2.8 |
| Ground thermal conductivity | $k_g$ | W/(m K) | 1.8 | 1.6 | 1.2 |
| Ground volumetric heat capacity | $(\rho c)_g$ | MJ/(m³K) | 2.8 | 3.0 | 2.2 |
| Volume flow rate | $\dot{V}$ | liters/min | 14 | 22 | 12 |
| Heat rate per unit length | $\dot{q}_l$ | W/m | 50 | 50 | 50 |

The mesh independence of the results for BHE 1 is illustrated in Figure 1, where the time evolutions of $T_{in}$, $T_{out}$, $T_{fm}$, and $T_b$ obtained with Mesh 3, having 575,860 elements, are compared with those obtained with Mesh 2, having 949,178 elements. The results are reported at 56 time instants, equally spaced in a logarithmic scale from $10^{-2.5}$ hours (about 11.4 s) to $10^3$ hours. The root mean square difference between the results obtained with the two meshes is 0.007 °C for $T_{in}$ and $T_{fm}$, 0.008 °C for $T_{out}$, and 0.002 °C for $T_b$. A very steep increase of $T_{in}$ and $T_{out}$ occurs after $\log_{10}(t_{hours}) = -0.7$, i.e. $t = 11.97$ minutes, namely 1.54 minutes after the completion of the circuit by the heated fluid. Before that instant, $T_{in}$ and $T_{out}$ remain constant.

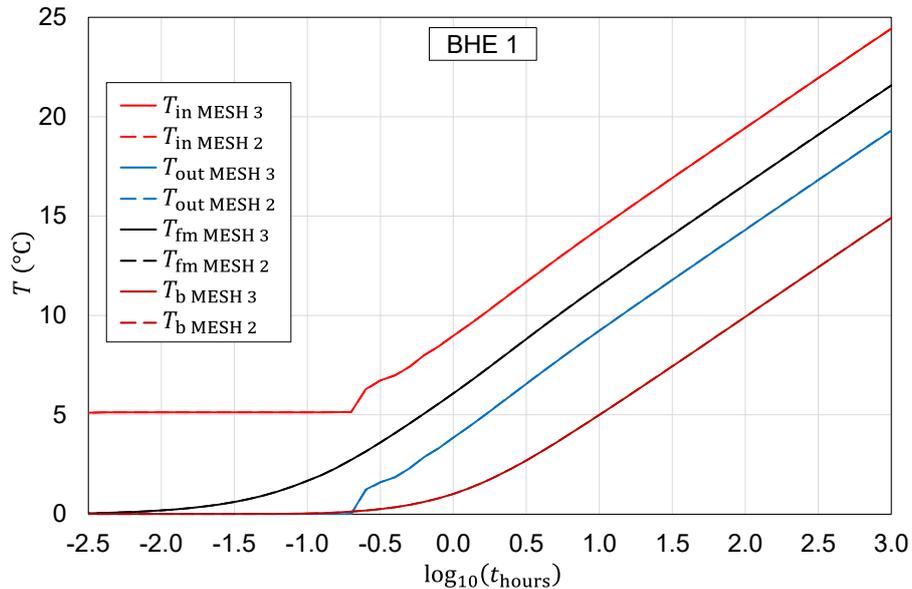

Figure 1: Time evolutions of $T_{in}$, $T_{out}$, $T_{fm}$ and $T_b$ for BHE 1, obtained by Mesh 2 and Mesh 3.

In order to determine the errors due to the HUST condition, we performed an additional simulation of BHE 1, where this condition is forced. In this simulation, a ground cylinder with thickness 4 mm, adjacent to the BHE surface, is replaced by a layer having thermal conductivity $10^6$ W/(m K) in the horizontal directions, zero in the vertical direction, and the same volumetric heat capacity as the ground. A thin resistive layer with thickness 0.1 mm and thermal resistance per unit length 0.004535 m K/W is inserted after the high-conductivity layer, to restore the real thermal resistance of the ground. The complete meshed domain and a particular of the mesh at the BHE top are illustrated in Figure 2. The mesh consists of 654,145 tetrahedral elements. The flat shape of the computational domain is due to the rescaling of the vertical coordinate.

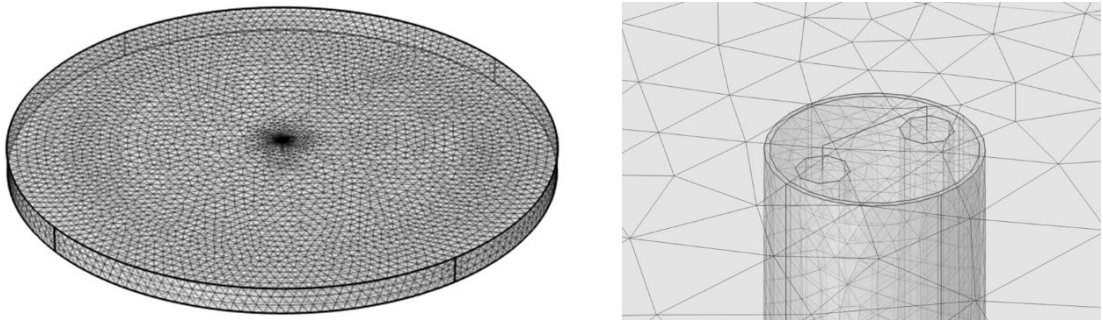

Figure 2: Mesh employed in the simulation of BHE 1 with the high-conductivity layer: complete meshed domain (left) and particular at the BHE top (right).

The time evolutions of $T_{in}$, $T_{out}$, $T_{fm}$, and $T_b$ obtained in the real conditions have been compared with those obtained in the simulation with the high-conductivity layer, as illustrated in Figure 3.

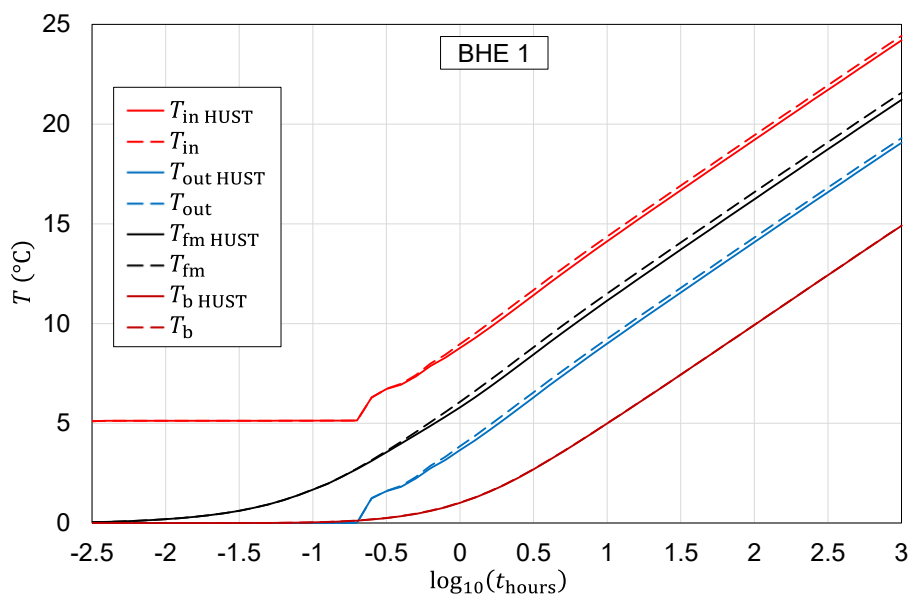

Figure 3: Time evolutions of $T_{in}$, $T_{out}$, $T_{fm}$, and $T_b$ for BHE 1, in the case of horizontally uniform surface temperature (HUST) of the BHE and in the real case.

The comparison shows that the HUST condition yields an underestimation of $T_{\text{in}}$, $T_{\text{out}}$, and $T_{\text{fm}}$, while it has no appreciable effect on $T_{\text{b}}$. The underestimation of $T_{\text{fm}}$ is slightly greater than that of $T_{\text{in}}$ and $T_{\text{out}}$.

The effect of the high-conductivity layer on $R_{\text{b3D}}$ and $R_{\text{beff}}$ is illustrated in Figure 4.

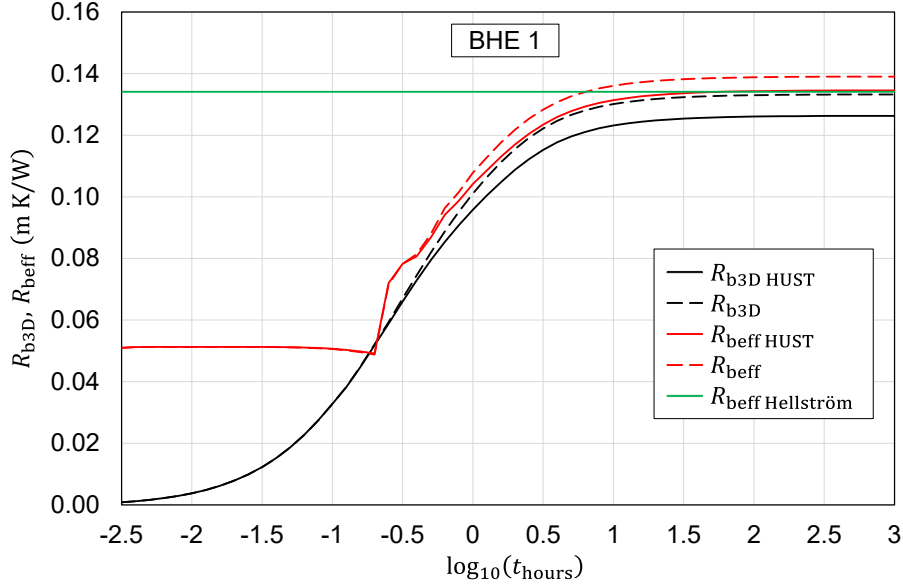

Figure 4: Time evolutions of $R_{\text{b3D}}$ and $R_{\text{eff}}$ for BHE 1, in the case of horizontally uniform surface temperature (HUST) of the BHE and in the real case. The green line is the effective borehole thermal resistance given by Hellström's equation.

The figure shows that the HUST condition yields an underestimation of $R_{\text{b3D}}$ and $R_{\text{beff}}$, and that the underestimation of $R_{\text{b3D}}$ is greater than that of $R_{\text{beff}}$. The asymptotic value of $R_{\text{beff}}$ obtained in the case of the HUST condition is nearly coincident with the stationary value of $R_{\text{beff}}$ given by Hellström's equation [22],

$$R_{\text{beff}} = \eta \coth(\eta) \, R_{\text{b}}, \tag{6}$$

where $\eta$ is the dimensionless parameter

$$\eta = \frac{L}{\dot{m}_f c_f} \frac{1}{\sqrt{R_a R_b}}, \tag{7}$$

and $R_{\text{a}}$ is the total thermal resistance per unit length between the descending and the ascending flow. In summary, the simulation with the high-conductivity layer shows that the HUST condition yields an underestimation of $T_{\text{in}}$, $T_{\text{out}}$, $T_{\text{fm}}$, $R_{\text{b3D}}$ and $R_{\text{beff}}$, while it has a negligible effect on $T_{\text{b}}$. This suggests that the HUST condition may have the same effect on the results of TRCMs.

## 3. Description of the proposed TRCM

In the proposed TRCM, the BHE and the ground are divided into $m$ horizontal slices of the same height $l$. Each slice is modeled with a thermal circuit based on the standard triangular network of thermal resistances [22], to which additional nodes are added.

The standard triangular network consists of three nodes: $f_1$, which represents the downward flow of the fluid, $f_2$, which represents the upward flow, and b, which represents the external surface of the BHE. The nodes are connected by the thermal resistances per unit length $R_1^\Delta$, $R_2^\Delta$, $R_{12}^\Delta$, as illustrated in Figure 5. Node $f_1$ exchanges heat with b through the thermal resistance $R_1^\Delta$, $f_2$ exchanges heat with b through the thermal resistance $R_2^\Delta$, and $f_1$ exchanges heat with $f_2$ through the thermal resistance $R_{12}^\Delta$.

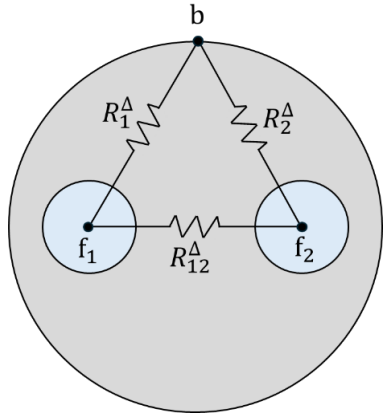

Figure 5: Standard triangular network of thermal resistances: $R_1^\Delta$ is the thermal resistance between fluid 1 and the borehole wall, $R_2^\Delta$ is the thermal resistance between fluid 2 and the borehole wall, and $R_{12}^\Delta$ is the thermal resistance between the flows.

This paper only deals with symmetric pipes, for which $R_2^\Delta = R_1^\Delta$. In this case, the thermal resistances $R_1^\Delta$ and $R_{12}^\Delta$ can be easily obtained from the borehole thermal resistance, $R_b$, and the total thermal resistance between the flows, $R_a$, by the following equations:

$$R_1^\Delta = 2R_b, \tag{8}$$

$$R_{12}^\Delta = \frac{4R_b R_a}{4R_b - R_a}. \tag{9}$$

The values of $R_b$ and $R_a$ used in our simulations have been obtained from the first-order multipole expressions given by Javes and Spitler [20].

In our model, several nodes are added to the standard triangular network. Four nodes are added to better describe the internal structure of the BHE: $p_1$, representing pipe 1, $p_2$, representing pipe 2, $gt_1$, which represents a portion of grout surrounding pipe 1, and $gt_2$, which represents a portion of grout surrounding pipe 2. The ground is divided into $n$ concentric annuli, each modeled with a node:

$g_1, \ldots, g_n$. Finally, one last node, $g_{n+1}$, is added to represent the unperturbed ground. A diagram of the thermal circuit is shown in Figure 6.

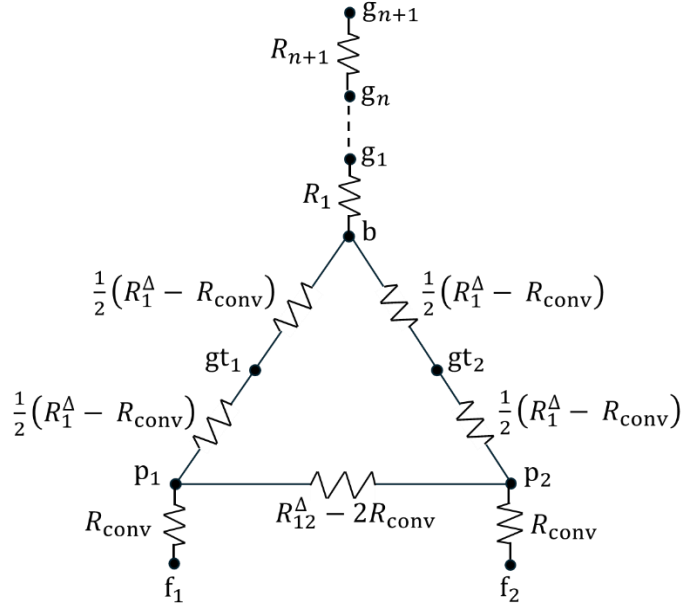

Figure 6: Network of thermal resistances of the proposed model, for each horizontal slice.

For simplicity, the mass of the pipes is assumed to be concentrated at their inner surface. Therefore, the thermal resistance between $f_1$ and $p_1$ (as well as the thermal resistance between $f_2$ and $p_2$) is

$$R_{\text{conv}} = \frac{1}{2\pi r_{\text{pi}} h}, \tag{10}$$

where $h$ is the convection coefficient and $r_{\text{pi}}$ the inner radius of the pipe.

The other thermal resistances inside the BHE can be determined by comparing our thermal circuit with that illustrated in Figure 5. The thermal resistance between $p_1$ and $p_2$ can be obtained by subtracting $2R_{\text{conv}}$ from $R_{12}^\Delta$. The thermal resistance between $p_1$ and b can be obtained by subtracting $R_{\text{conv}}$ from $R_1^\Delta$. We chose to split this resistance in half by introducing the node $gt_1$, and we did the same for the resistance between $p_2$ and b, by introducing the node $gt_2$.

As for the ground, let $r_0$ be the radius of the BHE and $r_i$ the outer radius of the $i$-th annulus. The thermal resistances $R_1, \ldots R_{n+1}$ appearing in Figure 2 are given by the equation:

$$R_i = \frac{1}{2\pi k_g} \ln\left(\frac{\bar{r}_i}{\bar{r}_{i-1}}\right), \tag{11}$$

where $k_g$ is the thermal conductivity of the ground and

$$\bar{r}_i = \begin{cases} r_0 & \text{for } i = 0 \\ \sqrt{(r_{i-1}^2 + r_i^2)/2} & \text{for } i = 1, \ldots n \\ r_n & \text{for } i = n+1 \end{cases} \tag{12}$$

The partition of the ground, in the simple case of two concentric annuli, is illustrated in Figure 7.

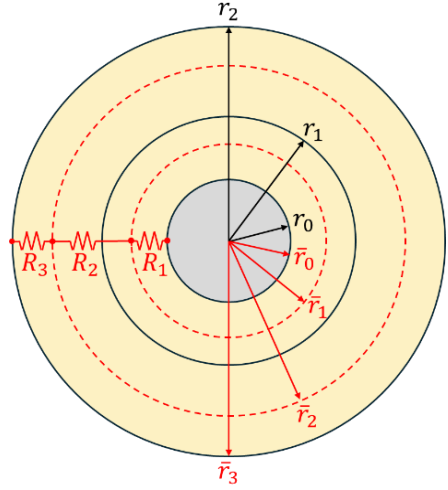

Figure 7: Partition of the ground in the case of two concentric annuli.

Note that $R_{\text{conv}}$ and $R_i$, as well as $R_1^\Delta$, $R_{12}^\Delta$, $R_a$ and $R_b$, are thermal resistances per unit length. The corresponding thermal resistances in $\text{K W}^{-1}$, for each horizontal layer of heigh $l$, are equal to the former divided by $l$.

Since each node models a portion of space filled with a given material, its heat capacity can be obtained by multiplying the volume of the portion of space, the density $\rho$ of the material and its specific heat capacity $c$. Thus, each ground node $g_i$ has heat capacity

$$C_i = \rho_g c_g \pi (r_i^2 - r_{i-1}^2) l, \tag{13}$$

each fluid node has heat capacity

$$C_f = \rho_f c_f \pi r_{\text{pi}}^2 l, \tag{14}$$

and each pipe node has heat capacity

$$C_p = \rho_p c_p \pi (r_{\text{pe}}^2 - r_{\text{pi}}^2) l, \tag{15}$$

where $r_{\text{pe}}$ is the outer radius of the pipe. As for the grout, a single layer of height $l$ has heat capacity

$$C_{\text{gt}} = \rho_{\text{gt}} c_{\text{gt}} \pi (r_0^2 - 2 r_{\text{pe}}^2) l. \tag{16}$$

This is distributed among the nodes $\text{gt}_1$, $\text{gt}_2$ and b. We can then call $(1-\varepsilon) C_{\text{gt}}$ the heat capacity of node b, and $(\varepsilon/2) C_{\text{gt}}$ the heat capacity of each of the nodes $\text{gt}_1$ and $\text{gt}_2$, where $\varepsilon$ is a number between 0 and 1.

Next, let us denote by a subscript $j$ the lower surface of the $j$-th layer. Let us call $T_{\text{f1},j}$ and $T_{\text{f2},j}$ the temperatures of the fluid nodes, $T_{\text{p1},j}$ and $T_{\text{p2},j}$ the temperatures of the pipe nodes, $T_{\text{gt1},j}$ and $T_{\text{gt2},j}$ the temperatures of the grout nodes, $T_{0,j}$ the temperature of the borehole node, and $T_{1,j}, \ldots, T_{n,j}$ the temperatures of the ground nodes, all measured at the lower surface of the of the $j$-th layer.

Since the two pipes of the BHE are connected at the bottom,

$$T_{f1,m} = T_{f2,m}. \tag{17}$$

Let us also call $T_{f1,0} = T_{in}$ the inlet fluid temperature and $T_{f2,0} = T_{out}$ the outlet fluid temperature. The energy balance equation yields

$$T_{f1,0} = T_{f2,0} + \frac{\dot{Q}}{\dot{m}_f c_f}, \tag{18}$$

where $\dot{m}_f$ is the mass flow rate of the fluid and $\dot{Q}$ is the heat rate supplied to the BHE.

Finally, let us set the undisturbed ground temperature equal to zero. Namely,

$$T_{n+1,j} = T_g = 0. \tag{19}$$

Now, let us consider a set of time instants $t_k$, with $k = 0, 1, \ldots, K$, where $t_0 = 0$ is the start of the heating. Assume that the temperatures of the nodes are constant for $t_k < t < t_{k+1}$, and that heat conduction in the vertical direction is negligible, so that energy transfer in the vertical direction takes place only through the fluid flow.

For every time instant (namely, for $k = 1, \ldots, K$) and for every layer of height $l$ (namely, for $j = 1, \ldots, m$), the following balance equations hold.

1. Energy balance for the fluid nodes:

$$\dot{m}_f c_f \left( T_{f1,j-1}(t_k) - T_{f1,j}(t_k) \right) - \frac{T_{f1,j}(t_k) - T_{p1,j}(t_k)}{R_{conv}/l} = C_f \frac{T_{f1,j}(t_k) - T_{f1,j}(t_{k-1})}{t_k - t_{k-1}}, \tag{20}$$

$$\dot{m}_f c_f \left( T_{f2,j}(t_k) - T_{f2,j-1}(t_k) \right) - \frac{T_{f2,j}(t_k) - T_{p2,j}(t_k)}{R_{conv}/l} = C_f \frac{T_{f2,j}(t_k) - T_{f2,j}(t_{k-1})}{t_k - t_{k-1}}. \tag{21}$$

2. Energy balance for the pipe nodes:

$$\frac{T_{f1,j}(t_k) - T_{p1,j}(t_k)}{R_{conv}/l} + \frac{T_{p2,j}(t_k) - T_{p1,j}(t_k)}{(R_{12}^\Delta - 2R_{conv})/l} + \frac{T_{gt1,j}(t_k) - T_{p1,j}(t_k)}{\frac{1}{2}(R_1^\Delta - R_{conv})/l} = C_p \frac{T_{p1,j}(t_k) - T_{p1,j}(t_{k-1})}{t_k - t_{k-1}}, \tag{22}$$

$$\frac{T_{f2,j}(t_k) - T_{p2,j}(t_k)}{R_{conv}/l} + \frac{T_{p1,j}(t_k) - T_{p2,j}(t_k)}{(R_{12}^\Delta - 2R_{conv})/l} + \frac{T_{gt2,j}(t_k) - T_{p2,j}(t_k)}{\frac{1}{2}(R_1^\Delta - R_{conv})/l} = C_p \frac{T_{p2,j}(t_k) - T_{p2,j}(t_{k-1})}{t_k - t_{k-1}}. \tag{23}$$

3. Energy balance for the grout nodes:

$$\frac{T_{p1,j}(t_k) - T_{gt1,j}(t_k)}{\frac{1}{2}(R_1^\Delta - R_{conv})/l} + \frac{T_{0,j}(t_k) - T_{gt1,j}(t_k)}{\frac{1}{2}(R_1^\Delta - R_{conv})/l} = \frac{\varepsilon}{2} C_{gt} \frac{T_{gt1,j}(t_k) - T_{gt1,j}(t_{k-1})}{t_k - t_{k-1}}. \tag{24}$$

$$\frac{T_{p2,j}(t_k) - T_{gt2,j}(t_k)}{\frac{1}{2}(R_1^\Delta - R_{conv})/l} + \frac{T_{0,j}(t_k) - T_{gt2,j}(t_k)}{\frac{1}{2}(R_1^\Delta - R_{conv})/l} = \frac{\varepsilon}{2} C_{gt} \frac{T_{gt2,j}(t_k) - T_{gt2,j}(t_{k-1})}{t_k - t_{k-1}}. \tag{25}$$

4. Energy balance for the borehole node:

$$\frac{T_{1,j}(t_k) - T_{0,j}(t_k)}{R_1/l} + \frac{T_{gt1,j}(t_k) - T_{0,j}(t_k)}{\frac{1}{2}(R_1^\Delta - R_{conv})/l} + \frac{T_{gt2,j}(t_k) - T_{0,j}(t_k)}{\frac{1}{2}(R_1^\Delta - R_{conv})/l} = (1 - \varepsilon) C_{gt} \frac{T_{0,j}(t_k) - T_{0,j}(t_{k-1})}{t_k - t_{k-1}}. \tag{26}$$

5. Energy balance for the ground nodes:

$$\frac{T_{i-1,j}(t_k) - T_{i,j}(t_k)}{R_i/l} + \frac{T_{i+1,j}(t_k) - T_{i,j}(t_k)}{R_{i+1}/l} = C_i \frac{T_{i,j}(t_k) - T_{i,j}(t_{k-1})}{t_k - t_{k-1}}, \tag{27}$$

with $i = 1, \ldots, n$. Eqs. (20-27), together with Eqs. (17) and (18), form a linear system of $(n + 7)j + 2$ equations in the unknowns $T_{f1,0}, T_{f2,0}, T_{f1,j}, T_{f2,j}, T_{p1,j}, T_{p2,j}, T_{gt1,j}, T_{gt2,j}, T_{0,j}, T_{1,j}, \ldots, T_{n,j}$. The system can be solved for increasing values of $k$. For $k = 0$, all the temperatures are zero, except the inlet fluid temperature, $T_{f1,0}(0)$, which is given by Eq. (18).

## 4. Comparison between the results of the TRCM and those of the finite-element model

As explained in Section 2, the comparison between the results of the finite-element simulations with and without the high-conductivity layer suggests that it is not possible to reproduce exactly the time evolutions of $R_{b3D}$ and of $T_{fm}$ by means of a TRCM that employs the HUST assumption. On the other hand, the comparison suggests that it is possible to get accurate values of $T_b$. For this purpose, we coded a C++ program that implements the TRCM illustrated in Section 3 with a selectable number of horizonal slices and a customizable partition of the ground. To obtain good accuracy, we performed our simulations with 100 horizontal slices and a finely partitioned ground: we set the thickness of the first ground annulus to 1 cm, the thickness increase ratio to 1.25, and the outer radius of the last ground annulus to be greater than 15 m, obtaining a total of 27 annuli. As for the value of $\varepsilon$, after performing many simulations we decided to fix it to 0.3.

The C++ program allows the user to change the geometrical and thermal parameters of the BHE, the number of horizontal layers and the partition of the ground. As output, the program yields the time evolutions of $T_{in}$, $T_{out}$, $T_{fm}$, $T_b$, $R_{beff}$ and $R_{b3D}$.

The time evolutions of $T_{in}$, $T_{out}$, $T_{fm}$, and $T_b$ obtained for BHE 1 using the C++ program are compared with those obtained by the finite-element model in Figure 8.

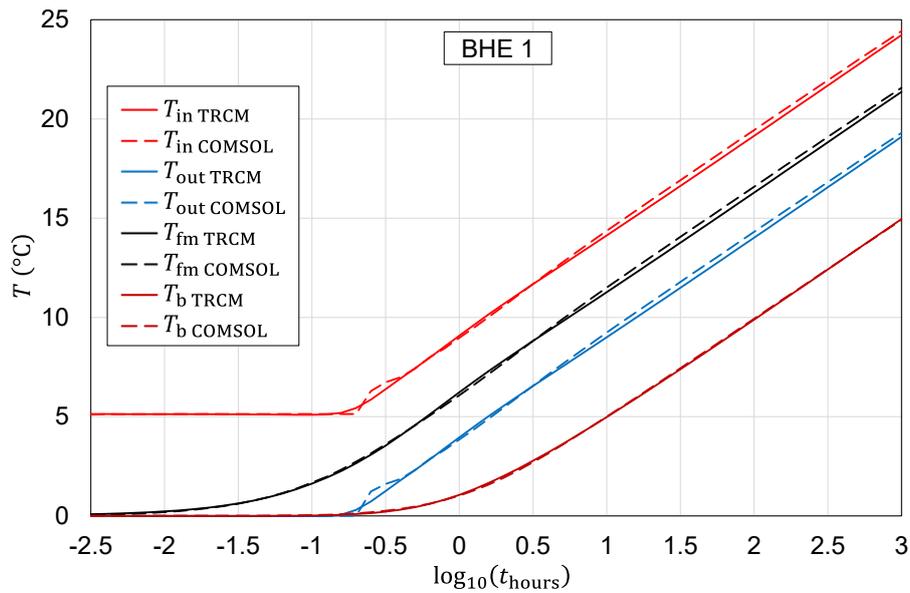

Figure 8: Time evolutions of $T_{in}$, $T_{out}$, $T_{fm}$ and $T_b$ obtained by the TRCM and by the finite-element model, for BHE 1.

The figure shows that the TRCM determines rather accurately the time evolution of $T_{in}$, $T_{out}$, and $T_{fm}$ up to $\log_{10}(t_{hours}) \approx 0.6$, i.e. $t \approx 4$ hours, but slightly underestimates these temperatures for greater values of time. On the other hand, the TRCM determines the time evolution of $T_b$ very accurately, with a root mean square difference of 0.037 °C with respect to the finite-element simulation.

As for $R_{b3D}$ and $R_{beff}$, the results obtained for BHE 1 using the C++ program are compared with those obtained by the finite-element model in Figure 9. The figure shows that the TRCM underestimates $R_{b3D}$ and $R_{beff}$ for $\log_{10}(t_{hours}) > 0.5$, and gives an asymptotic value of $R_{beff}$ equal to the stationary value given by Hellström's equation, which is obtained under the assumption of isothermal surface of the BHE.

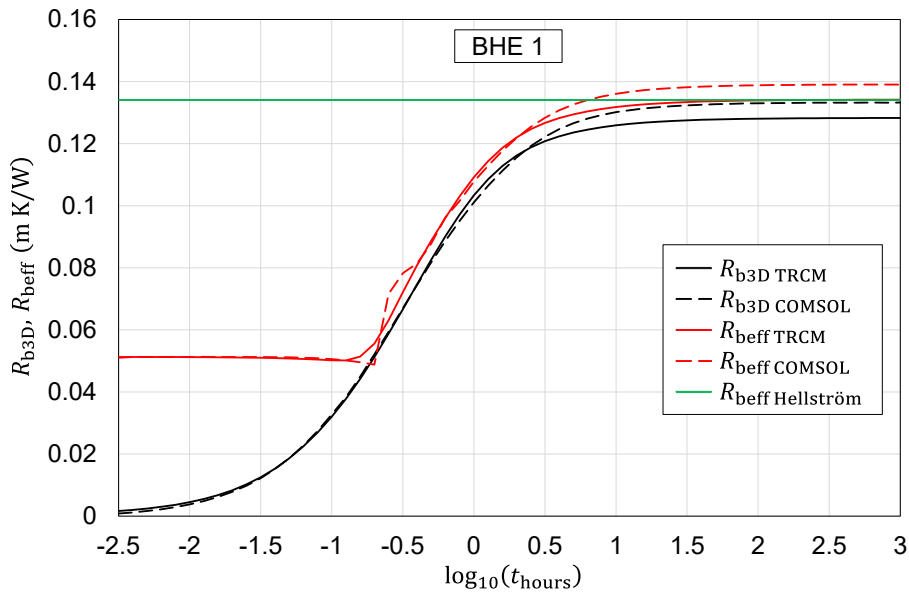

Figure 9: Time evolutions of $R_{b3D}$ and $R_{beff}$ obtained by the TRCM and by the finite-element model, for BHE 1. The green line is the effective borehole thermal resistance given by Hellström's equation.

Finally, the time evolution of $T_{fave} - T_{fm}$ obtained for BHE 1 using the C++ program is compared with that obtained by the finite-element model in Figure 10. The figure shows that the TRCM reproduces the time evolution of $T_{fave} - T_{fm}$ accurately, except in the range $-0.8 \leq \log_{10}(t_{hours}) \leq 0.5$, where the time evolution determined by COMSOL has an irregular trend. Analogous results to those obtained for BHE 1 have been found for all the BHEs studied in this paper.

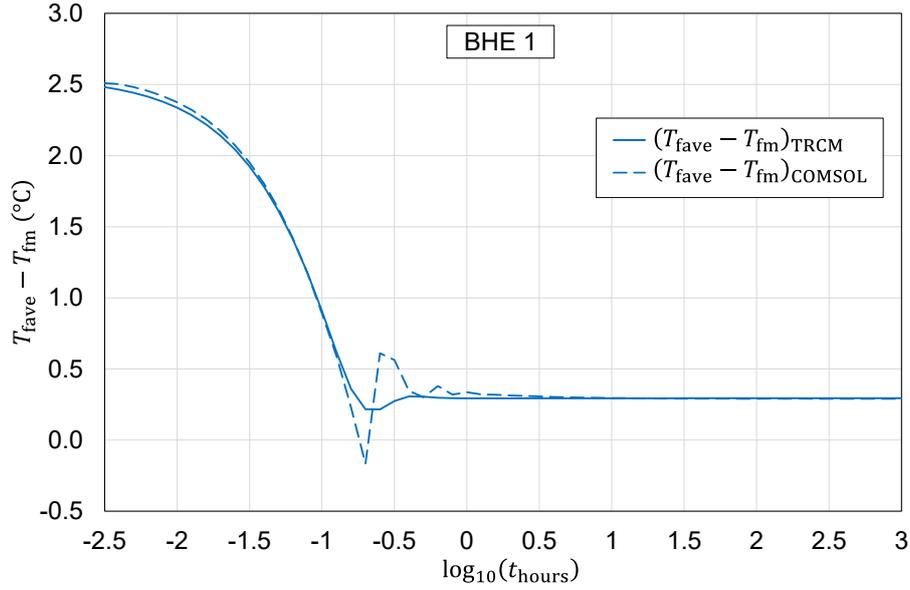

Figure 10: Time evolution of $T_{\text{fave}} - T_{\text{fm}}$ obtained by the TRCM and by the finite-element model, for BHE 1.

In summary, the proposed TRCM determines correctly the time evolutions of $T_{\text{b}}$ and of $T_{\text{fave}} - T_{\text{fm}}$, but slightly underestimates the time evolutions of $T_{\text{in}}$, $T_{\text{out}}$, $T_{\text{fm}}$, $R_{\text{b3D}}$ and $R_{\text{beff}}$ in the medium term. The underestimation is similar to the one observed for the finite-element simulation with the forced HUST condition. This reinforces the hypothesis that employing the HUST assumption causes unavoidable inaccuracies in the results of TRCMs.

To confirm this hypothesis, we compared the results of our simulations with those of the ANN developed by Pasquier and Marcotte [43]. The latter is based on the TRCM by Pasquier and Marcotte [35], which employs the HUST assumption. The MATLAB code of the ANN was kindly supplied by Pasquier.

For the comparison, a few changes were made to BHE 1. The volumetric heat capacity of the pipe was set to 1.9 MJ/(m³K) and the volume flow rate of the fluid to 16 liters per minute, since our original values were out of the range allowed by the ANN. Moreover, the convective thermal resistance per unit length between fluid and pipes was set to $10^{-6}$ m K/W, as in the ANN. The heat rate $\dot{q}_{\text{l}} = 50$ W/m was imposed in the ANN by setting $T_{\text{in}} - T_{\text{out}} = \frac{\dot{q}_{\text{l}} L}{\dot{V}(\rho c)_{\text{f}}} = 4.4893$ °C. The remaining input parameters are those listed in Table 1, fourth column.

The simulation of BHE 1, with the changes listed above, was performed with the ANN [43], with our TRCM and with COMSOL Multiphysics. A comparison of the time evolutions of $T_{\text{out}}$ is illustrated in Figure 11.

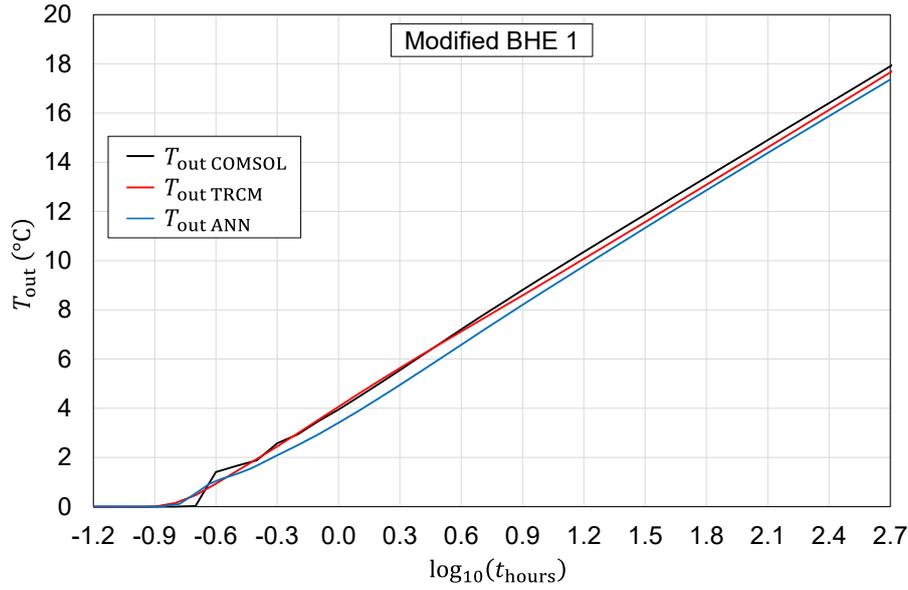

Figure 11: Time evolution of $T_{out}$ determined by COMSOL Multiphysics, by our TRCM, and by the ANN developed by Pasquier and Marcotte [43]. The results refer to a modified BHE 1, with $(\rho c)_p = 1.9 \text{ MJ}/(\text{m}^3\text{K})$, $\dot{V} = 16$ liters per minute and $R_{conv} = 10^{-6}$ m K/W.

As expected, both our TRCM and the ANN [43] underestimate $T_{out}$ for $\log_{10}(t_{hours}) > 0.6$, and the underestimation by the ANN is slightly larger.

The results of this section confirm that the HUST assumption causes an inevitable underestimation of the medium-term thermal response in TRCMs. This assumption, however, is necessary to use the correct expressions of the thermal resistances $R_b$ and $R_a$. Therefore, the only way to obtain accurate time evolutions of $T_{in}$, $T_{out}$, $T_{fm}$, $R_{b3D}$, and $R_{beff}$ by a TRCM, even in the medium term, is to introduce a correction factor a posteriori.

## 5. Improvement of the TRCM model by a correction factor

Since our TRCM predicts accurately the time evolution of $T_b$, we can deduce from Eq. (1) that the underestimation of $T_{fm}$ is only due to that of $R_{b3D}$. Moreover, since our TRCM predicts accurately the time evolution of $T_{fave} - T_{fm}$, we can conclude that the underestimation of $T_{fave}$ is almost the same as that of $T_{fm}$. Therefore, by correcting the time evolution of $R_{b3D}$, it is possible to get accurate values of both $T_{fm}$ and $T_{fave}$. Finally, the corrected values of $T_{fave}$ and the energy balance equation can be used to obtain accurate values of $T_{in}$ and $T_{out}$, while the corrected values of $R_{beff}$ can be obtained from Eq. (4).

In short, the inaccuracies of the TRCM can be nearly completely resolved by determining, for each U-tube BHE, a time-dependent correction coefficient that transforms the time evolution of $R_{b3D}$ given by the TRCM into that given by the finite-element simulation.

We define the correction coefficient as

$$c_{\text{coeff}} = \frac{R_{\text{b3D,COMSOL}}}{R_{\text{b3D,TRCM}}}. \tag{28}$$

This way, the product $c_{\text{coeff}} R_{\text{b3D,TRCM}}$ yields a corrected value of the 3D borehole thermal resistance, which we will call $R_{\text{b3D,c}}$, equal to the one determined by COMSOL.

Once the correction coefficient is known, the corrected values of the mean, the average, the inlet and the outlet fluid temperature, $T_{\text{fm,c}}$, $T_{\text{fave,c}}$, $T_{\text{in,c}}$, $T_{\text{out,c}}$, and the corrected values of the effective borehole thermal resistance, $R_{\text{beff,c}}$, can then be determined as follows:

$$T_{\text{fm,c}} = T_{\text{b}} + \dot{q}_l R_{\text{b3D,c}}, \tag{29}$$

$$T_{\text{fave,c}} = T_{\text{fm,c}} + (T_{\text{fave}} - T_{\text{fm}}), \tag{30}$$

$$T_{\text{in,c}} = T_{\text{fave,c}} + 0.5 \frac{\dot{Q}}{\dot{m} c_{\text{f}}}, \tag{31}$$

$$T_{\text{out,c}} = T_{\text{fave,c}} - 0.5 \frac{\dot{Q}}{\dot{m} c_{\text{f}}}. \tag{32}$$

$$R_{\text{beff,c}} = \frac{1}{\dot{q}_l} (T_{\text{fave,c}} - T_{\text{b}}). \tag{33}$$

Since $T_{\text{b}}$ is already estimated accurately, there is no need to correct it.

Several pairs of simulations, performed by COMSOL Multiphysics and by the TRCM, showed that only four parameters have a relevant effect on the correction coefficient: the borehole radius, $r_{\text{b}}$, the shank spacing, $s$, the thermal conductivity of the grout, $k_{\text{gt}}$, and the outer radius of the pipe, $r_{\text{pe}}$. Since single U-tube BHEs usually have standard pipes, with $r_{\text{pe}} = 20$ mm or $r_{\text{pe}} = 16$ mm, we selected these two values. For each value of $r_{\text{pe}}$, we considered three values of $r_{\text{b}}$, namely 68, 72 and 76 mm, three values of $s$, namely 27, 37 and 47 mm, and three values of $k_{\text{gt}}$, namely 1.0, 1.6 and 2.2 W/(m K). For the BHEs with $r_{\text{pe}} = 20$ mm, we set $r_{\text{pi}} = 16.3$ mm and $\dot{V}_{\text{f}} = 14$ liters per minute; for the BHEs with $r_{\text{pe}} = 16$ mm, we set $r_{\text{pi}} = 13$ mm and $\dot{V}_{\text{f}} = 12$ liters per minute. The selected values of $r_{\text{pi}}$ are those corresponding to $r_{\text{pe}}$ in the standard pipes. We kept fixed all the remaining parameters, with the following values: $L = 100$ m, $(\rho c)_{\text{gt}} = 2.5$ MJ/(m³K), $k_{\text{g}} = 1.8$ W/(m K), $(\rho c)_{\text{g}} = 2.8$ MJ/(m³K), and $D$, $k_{\text{p}}$, $(\rho c)_{\text{p}}$, $\dot{q}_l$ with the values reported in Table 1. We selected water as the working fluid, with properties evaluated at 20 °C. In total, we ran 54 pairs of simulations with COMSOL Multiphysics and with the TRCM, 27 for each value of $r_{\text{pe}}$, obtaining 54 time-dependent values of $c_{\text{coeff}}$. The $R_{\text{b3D}}$ values determined by COMSOL and the correction coefficients have been collected in an Excel file, available at the data repository of the University of Bologna.

A run option has been implemented in the C++ program, where the 54 time-dependent values of $c_{\text{coeff}}$ are used to correct the values of $T_{\text{in}}$, $T_{\text{out}}$, $T_{\text{fm}}$, $R_{\text{beff}}$, and $R_{\text{b3D}}$. If this option is chosen, parabolic interpolations are performed to estimate the correction factor for the values of $r_b$, $s$, and $k_{gt}$ set in the input file. A linear interpolation in $r_{pe}$ is also performed in case its value is neither 20 nor 16 mm.

## 6. Results

The time evolutions of $T_{in}$, $T_{out}$, $T_{fm}$, $T_b$, $R_{b3D}$ and $R_{beff}$ obtained by the corrected TRCM have been compared with those determined by the finite-element model for several single U-tube BHEs. An excellent agreement was found in all cases, except for small differences in the time evolutions of $T_{in}$, $T_{out}$, and $R_{beff}$ in the time ranges where the values given by COMSOL Multiphysics have an irregular trend. As examples, three BHEs are considered in this section: BHE 1, BHE 2 and BHE 3, whose geometrical and thermal parameters are reported in Table 1.

In the case of BHE 1, a particularly good agreement is expected between the finite-element model and the corrected TRCM. Indeed, BHE 1 is one of the 54 BHEs used to evaluate the correction coefficients, so no interpolation is required to apply the correction factor.

The time evolutions of $T_{in}$, $T_{out}$, and $T_{fm}$ obtained for BHE 1 by the corrected TRCM are compared with those determined by the finite-element model in Figure 12.

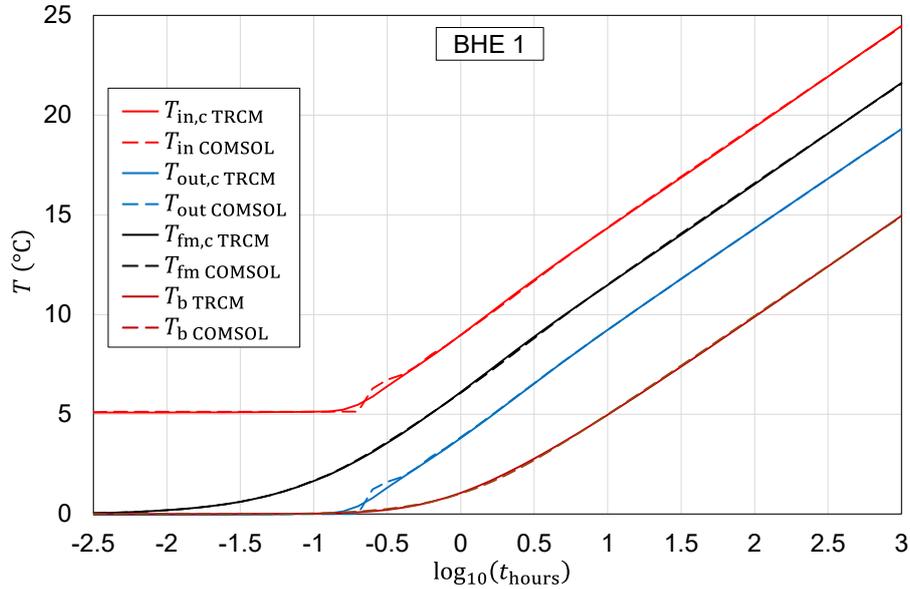

Figure 12: Time evolutions of $T_{in}$, $T_{out}$, $T_{fm}$ and $T_b$ obtained by the corrected TRCM and by the finite-element simulation run with COMSOL Multiphysics, for BHE 1.

The figure reveals an excellent agreement between the two models. The root mean square difference between the values of $T_{fm}$ given by the two simulations is 0.037 °C. The root mean square difference between the values of $T_b$ is also 0.037 °C, as reported in Section 4. As for $T_{in}$ and $T_{out}$, the values given by the corrected TRCM in time range $-0.8 \leq \log_{10}(t_{hours}) \leq -0.4$ are smoothed compared to those determined by COMSOL. Excluding this time range, the agreement between the two pairs of curves is excellent, with a root mean square difference of 0.038 °C for $T_{in}$ and 0.025 °C for $T_{out}$. The time evolutions of $R_{b3D}$ and $R_{beff}$ for BHE 1 are illustrated in Figure 13. Since BHE 1 is one the 54 BHEs used to evaluate the correction factors, the values of $R_{b3D}$ given by the two simulations are identical. Similarly to what we observed for $T_{in}$ and $T_{out}$, the values of $R_{beff}$ given by the corrected TRCM in time range $-0.8 \leq \log_{10}(t_{hours}) \leq -0.4$ are smoothed compared to those determined by COMSOL. Excluding this time range, the root mean square difference between the $R_{beff}$ values is $4.7 \times 10^{-4}$ m K/W. At the final instant $t = 1000$ hours, the value of $R_{beff}$ given by the corrected TRCM is $9.2 \times 10^{-5}$ m K/W greater than the one determined by COMSOL (namely, 0.066% greater).

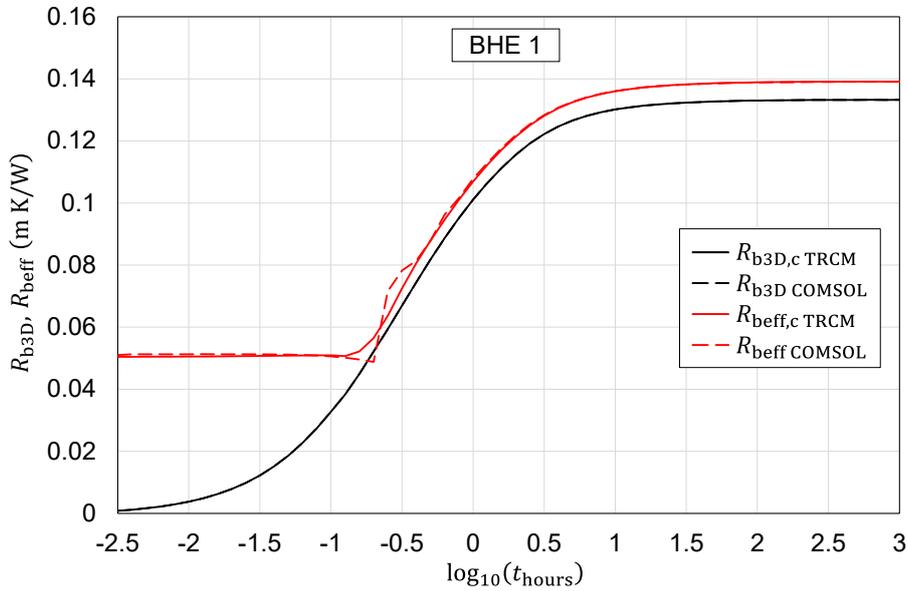

Figure 13: Time evolutions of $R_{b3D}$ and $R_{beff}$ obtained by the corrected TRCM and by the finite-element simulation run with COMSOL Multiphysics, for BHE 1.

Unlike BHE 1, almost all the parameters of BHE 2 are different from those used to determine the correction coefficients. In this case, the correction factor applied to the TRCM is estimated by the interpolation techniques mentioned in Section 5.

The time evolutions of $T_{in}$, $T_{out}$, $T_{fm}$ and $T_b$ obtained for BHE 2 by the corrected TRCM are compared with those determined by the finite-element model in Figure 14. The agreement between the two models is still excellent. The root mean square difference between the values of $T_{fm}$ by the two simulations is 0.051 °C. The root mean square difference between the values of $T_b$ is 0.047 °C. As for $T_{in}$ and $T_{out}$, the values given by the corrected TRCM in time range $-0.8 \leq \log_{10}(t_{hours}) \leq -0.3$ are smoothed compared to those determined by COMSOL. Excluding this time range, the agreement between the two pairs of curves is excellent, with a root mean square difference of 0.048 °C for $T_{in}$ and 0.049 °C for $T_{out}$.

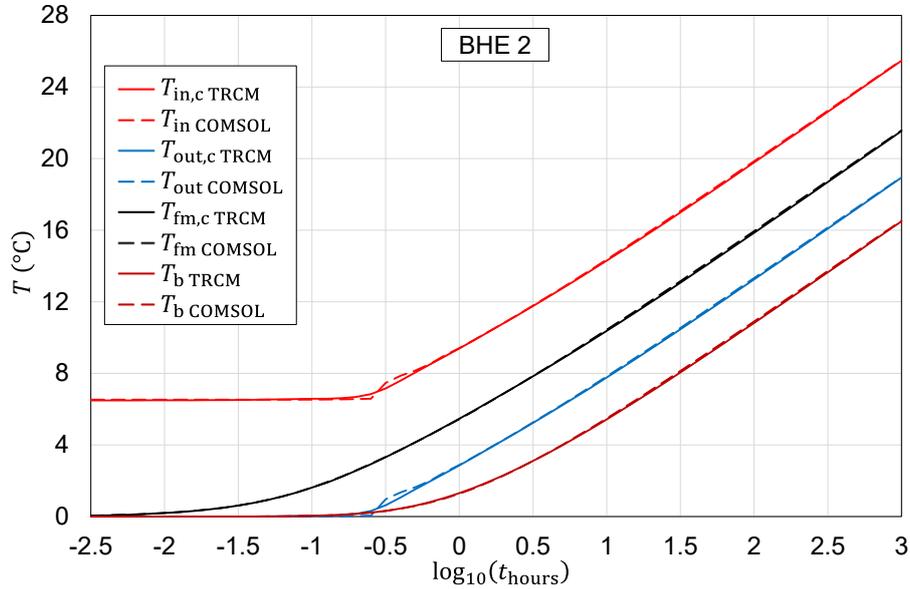

Figure 14: Time evolutions of $T_{in}$, $T_{out}$, and $T_{fm}$ and $T_b$ obtained by the corrected TRCM and by the finite-element simulation run with COMSOL Multiphysics, for BHE 2.

The time evolutions of $R_{b3D}$ and $R_{beff}$ for BHE 2 are illustrated in Figure 15. The root mean square difference between the $R_{b3D}$ values of the two simulations is $1.82 \times 10^{-4}$ m K/W. At $t = 1000$ hours, the value of $R_{b3D}$ given by the corrected TRCM is $1.54 \times 10^{-4}$ m K/W lower than the one determined by COMSOL (namely, 0.13% lower). The values of $R_{beff}$ given by the corrected TRCM in time range $-1 \leq \log_{10}(t_{hours}) \leq 0$ are smoothed compared to those determined by COMSOL. Excluding this time range, the root mean square difference between the $R_{beff}$ values is $4.61 \times 10^{-4}$ m K/W. At $t = 1000$ hours, the value of $R_{beff}$ given by the corrected TRCM is $2.59 \times 10^{-4}$ m K/W greater than the one determined by COMSOL (namely, 0.23% greater).

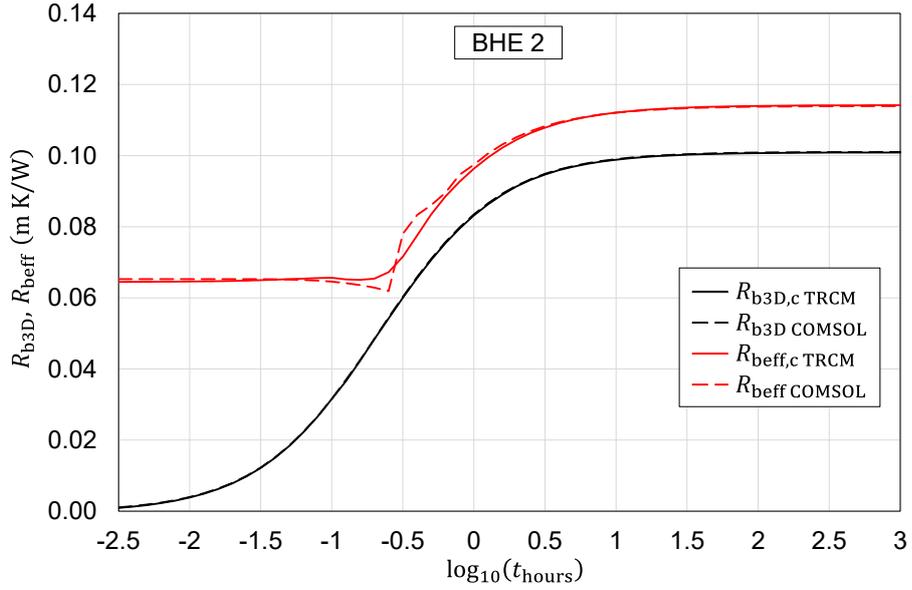

Figure 15: Time evolutions of $R_{b3D}$ and $R_{beff}$ obtained by the corrected TRCM and by the finite-element simulation run with COMSOL Multiphysics, for BHE 2.

As for BHE 3, not only do its parameters differ from those used to determine the correction coefficients, but the borehole radius, $r_b = 64$ mm, is smaller than the lowest value employed in the parabolic interpolation. This value of $r_b$ was chosen on purpose, to test the accuracy of the corrected TRCM in cases where extrapolation is required.

The time evolutions of $T_{in}$, $T_{out}$, $T_{fm}$ and $T_b$ obtained for BHE 3 by the corrected TRCM are compared with those determined by the finite-element model in Figure 16.

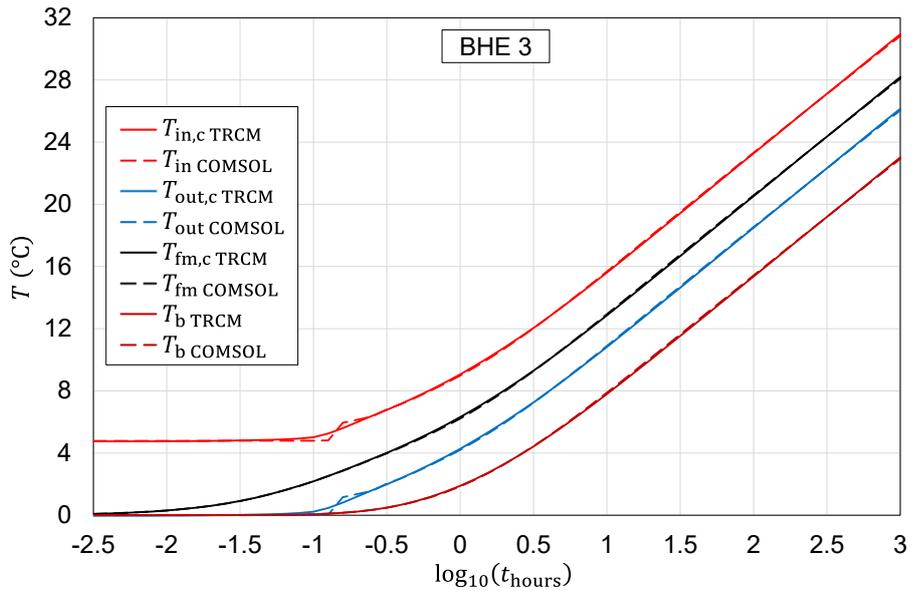

Figure 16: Time evolutions of $T_{in}$, $T_{out}$, $T_{fm}$ and $T_b$ obtained by the corrected TRCM and by the finite-element simulation run with COMSOL Multiphysics, for BHE 3.

The agreement between the two models is, once again, remarkable. The root mean square difference between the values of $T_{\text{fm}}$ given by the two simulations is $0.052$ °C. The root mean square difference between the values of $T_{\text{b}}$ is $0.049$ °C. As for $T_{\text{in}}$ and $T_{\text{out}}$, the values given by the corrected TRCM in time range $-1.2 \leq \log_{10}(t_{\text{hours}}) \leq -0.7$ are smoothed compared to those determined by COMSOL. Excluding this time range, the agreement between the two pairs of curves is excellent, with a root mean square difference of $0.052$ °C both for $T_{\text{in}}$ and $T_{\text{out}}$.

The time evolutions of $R_{\text{b3D}}$ and $R_{\text{beff}}$ for BHE 3 are illustrated in Figure 17. The root mean square difference between the $R_{\text{b3D}}$ values of the two simulations is $4.40 \times 10^{-4}$ m K/W. At $t = 1000$ hours, the value of $R_{\text{b3D}}$ given by the corrected TRCM is $2.10 \times 10^{-4}$ m K/W greater than the one determined by COMSOL (namely, 0.20% greater). The values of $R_{\text{beff}}$ given by the corrected TRCM in time range $-1.5 \leq \log_{10}(t_{\text{hours}}) \leq -0.5$ are smoothed compared to those determined by COMSOL. Excluding this time range, the root mean square difference between the $R_{\text{beff}}$ values is $5.26 \times 10^{-4}$ m K/W. At $t = 1000$ hours, the value of $R_{\text{beff}}$ given by the corrected TRCM is $4.43 \times 10^{-4}$ m K/W greater than the one determined by COMSOL (namely, 0.40% greater).

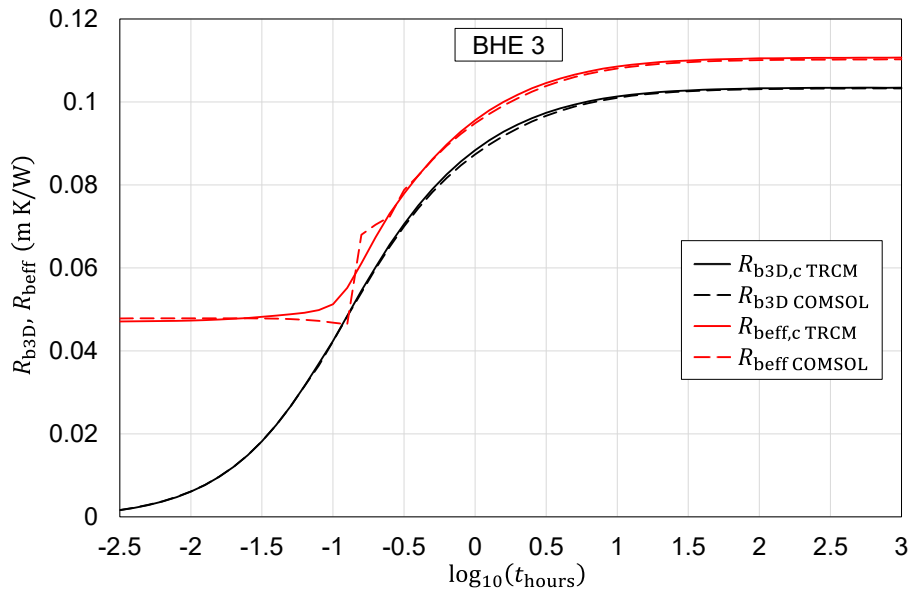

Figure 17: Time evolutions of $R_{\text{b3D}}$ and $R_{\text{beff}}$ obtained by the corrected TRCM and by the finite-element simulation run with COMSOL Multiphysics, for BHE 3.

## 7. Conclusions

In Thermal Resistance Capacity Models (TRCMs) of borehole heat exchangers (BHEs), the borehole wall is usually considered as horizontally isothermal, and is represented by one node at each depth. This is necessary in order to use the accurate expressions of the thermal resistances between each pipe and the borehole wall, and between the pipes, calculated by the multipole method.

We have shown that this assumption yields slight inaccuracies in the prediction of the time evolution of the inlet and outlet fluid temperatures, $T_{in}$ and $T_{out}$, of the mean fluid temperature, $T_{fm}$, of the 3D borehole thermal resistance, $R_{b3D}$, and of the effective borehole thermal resistance, $R_{beff}$, while it has no considerable effect on the time evolution of the mean surface temperature of the BHE, $T_b$.

This conclusion was reached by comparing two accurate finite-element simulations of a single U-tube BHE subjected to a constant heat rate, with and without a thin high-conductivity layer, adjacent to the BHE surface, that forces this surface to be horizontally isothermal. The simulations were implemented in COMSOL Multiphysics, and the energy balance along the flow was simulated through the Pipe Flow Module.

A TRCM of single U-tube BHEs has been developed. The model, which assumes a horizontally isothermal BHE surface, confirms the results obtained by the finite-element simulations: it slightly underestimates $T_{in}$, $T_{out}$, $T_{fm}$, $R_{b3D}$ and $R_{beff}$, but yields accurate time evolutions of $T_b$. The model also yields an accurate time evolution of the difference $T_{fm} - T_{ave}$, where $T_{ave}$ is the arithmetic mean of the inlet and outlet temperatures.

The accurate values of $T_b$ and $T_{fm} - T_{ave}$ have been exploited to design a method that corrects the results of the TRCM a posteriori. To this purpose, 54 pairs of simulations were performed, with our TRCM and with COMSOL Multiphysics, for different types of single U-tube BHEs. The results of the simulations were used to determine a time-dependent correction coefficient, $c_{coeff}$, that transforms the time evolution of $R_{b3D}$ calculated by our TRCM into that given by the COMSOL. The accurate time evolutions of $T_b$ and $T_{ave} - T_{fm}$, together with the corrected values of $R_{b3D}$, were used to obtain, by simple mathematical relations, accurate time evolutions of $T_{in}$, $T_{out}$, $T_{fm}$, and $R_{beff}$.

The proposed TRCM and the correction method have been implemented in a C++ program, available at the open-source online data repository of the University of Bologna. The program yields, within two seconds, the same time evolutions of $T_b$, $T_{in}$, $T_{out}$, $T_{fm}$, $R_{b3D}$, and $R_{eff}$ as those yielded by an accurate finite-element simulation requiring several hours of computation time. The results hold for any single U-tube BHE, in a working period from $10^{-2.5}$ hours (about 11 s) to $10^3$ hours.

The program can be used for very fast and accurate short- and medium-term simulations of single U-tube BHEs, or to provide accurate time dependent values of $R_{b3D}$ or $R_{beff}$, to be employed in full-time-scale simulations of bore fields.

**Acknowledgements**

The authors are grateful to Professor Philippe Pasquier for providing the MATLAB code of the ANN developed in Pasquier and Marcotte [43], which was used for comparison with our TRCM.